\documentclass[a4paper,fleqn,usenatbib]{mnras}

\usepackage[T1]{fontenc}

\usepackage{graphicx}	
\usepackage{amsmath}	
\usepackage{amssymb,amstext}	
\usepackage{bm}

\newcommand{\md}{\mathrm{d}}

\newcommand{\nn}{\nonumber}

\newcommand{\pder}[2]{\frac{\partial#1}{\partial#2}}

\newcommand{\bOm}{\mathbf{\Omega}}
\newcommand{\bk}{\mathbf{k}}

\newcommand{\bJ}{\mathbf{J}}

\newcommand{\pattern}{\Omega_\mathrm{p}}

 \title[Saturation of spiral instabilities]{Saturation of spiral instabilities in disk galaxies}

\author[C. Hamilton]{
  Chris Hamilton\thanks{E-mail: chamilton@ias.edu}\\
Institute for Advanced Study, Einstein Drive, Princeton, NJ 08540, USA}

\pubyear{2023}

\begin{document}
\label{firstpage}
\pagerange{\pageref{firstpage}--\pageref{lastpage}}
\maketitle


\begin{abstract}
Spiral density waves can arise in galactic disks as linear instabilities of the underlying stellar distribution function.  Such an instability grows exponentially in amplitude at some fixed growth rate $\beta$ before saturating nonlinearly. 
However, the mechanisms behind saturation, and the resulting saturated spiral amplitude, have received little attention.  Here we argue that one important saturation mechanism is the nonlinear trapping of stars near the spiral's corotation resonance. Under this mechanism, we show analytically that an $m$-armed spiral instability will saturate when the libration frequency of resonantly trapped orbits reaches $\omega_\mathrm{lib} \sim \mathrm{a\,\, few}\times  m^{1/2} \beta$. For a galaxy with a flat rotation curve this implies a maximum relative spiral surface density $\vert \delta\Sigma/\Sigma_0\vert \sim \mathrm{a\,\,few} \times (\beta/\Omega_\mathrm{p})^2 \cot \alpha$, where $\Omega_\mathrm{p}$ is the spiral pattern speed and $\alpha$ is its pitch angle. This result is in reasonable agreement with recent $N$-body simulations, and suggests that spirals driven by internally-generated instabilities reach relative amplitudes of at most a few tens of percent; higher amplitude spirals, like in M51 and NGC 1300, are likely caused by very strong bars and/or tidal perturbations.
\end{abstract}




\section{Introduction}
\label{sec:Introduction}

After decades of work there is still no consensus on the origin(s) and character of spiral structure in galaxies. Opinions differ over the mechanisms that provoke spiral responses, the lifetimes of individual spiral patterns, the importance of gas flows and star formation,
and so on --- for an array of perspectives, see e.g. reviews by \cite{athanassoula1984spiral,dobbs2014dawes,sellwood2022spirals}.
However, one area in which there are some robust results --- and which we hope 
bears some relation to real galaxies --- is the study of isolated, razor-thin, gas-free stellar disks.  In this highly simplified context, both $N$-body simulations and linear perturbation theory have demonstrated clearly that spiral structure can arise as a linear instability (i.e. an exponentially growing, rigidly rotating Landau mode) of the underlying axisymmetric stellar distribution function (DF) \citep{sellwood1991spiral, jalali2005unstable}.  For example, a `grooved' disk, in which there is a deficit of stars near a particular galactocentric radius (or more precisely, a localized deficit in the DF of orbital angular momenta), can be linearly unstable to spiral modes which exponentiate out of the finite-$N$ noise
\citep{Sellwood2014-xo,De_Rijcke2019-uo}.

The basic premise of the theory promoted by Sellwood \& Carlberg over recent decades --- and initiated by \cite{sellwood1989recurrent} --- is that once a spiral mode  has exponentiated to a large enough amplitude, it will resonantly scatter stars at its associated Lindblad resonances.  
This resonant scattering both drains the amplitude of the mode \textit{and} changes the underlying axisymmetric DF (e.g. by carving a new groove, see \citealt{Sellwood2019-gb}). 
The newly-modified axisymmetric DF is itself unstable to a new set of spiral modes, which scatter stars at their own Lindblad resonances, and so on in a `recurrent cycle of groove modes' (see also \citealt{dekker1976spiral,Sridhar2019-zo}).  

On the other hand, a key observable associated with spirals is their amplitude.
Observational studies of spiral amplitudes have historically suffered from small sample sizes, with at most $\sim 100$ galaxies per study and often much less --- see e.g. \cite{rix1995non,grosbol2004spiral,block2004gravitational,elmegreen2011grand}.  But this situation has changed dramatically in the past decade or so due to large-scale spectroscopic surveys like SDSS \citep{york2000sloan,strauss2002spectroscopic}, combined with classification efforts from 
citizen science \citep{lintott2008galaxy} and machine learning \citep{dominguez2018improving}, which facilitate statistical studies many thousands of galaxies \citep{masters2019galaxy,savchenko2020multiwavelength,porter2022galaxy,yu2022strong}.
Recent analyses have found that in galaxies without close companions, observable spiral amplitudes range from a few percent to several tens of percent as a fraction of the axisymmetric background; these amplitudes are most strongly (inversely) correlated with the central concentration of the galaxy, and correlate only weakly with other parameters like bar strength, pitch angle, star formation rate, and environmental characteristics \citep{yu2020statistical,smith2022effect}. 
Arms with relative overdensities $\gtrsim 100\%$ seem to exist only in galaxies that are undergoing strong tidal interactions like in M51,
or in very strongly barred galaxies like NGC 1300 \citep{elmegreen1989spiral,rix1993tracing,kendall2011spiral,kendall2015spiral}.

With regard to this observable, the groove-cycle paradigm is incomplete, since it does not answer the question: what sets the maximum amplitude of the recurrent spirals? The answer \textit{cannot} simply be that the scattering of stars at Lindblad resonances drains energy and angular momentum from the spiral modes --- in fact, \cite{sellwood2022spiral} showed that the mode amplitudes saturate before any significant changes to the DF near Lindblad resonances has occured\footnote{More precisely, these authors investigated saturation of spiral instabilties in $N$-body simulations of their aforementioned grooved razor-thin disks, mostly restricting the sectoral harmonics of the non-axisymmetric gravitational field to azimuthal waveumbers $m=2$ or $m=3$. See also \cite{sellwood1989recurrent, donner1994structure}.}.
Thus, some other process must be halting the exponential growth.
One natural candidate is the nonlinear trapping of stars at the spiral's corotation resonance,
which transports stars back and forth in angular momentum across the resonance, while leaving their radial actions unchanged.
\cite{sellwood2022spiral} showed that the saturation amplitude of the surface density perturbation scaled as 
$\vert \delta \Sigma \vert \propto \beta^2$, where $\beta$ is the linear instability's growth rate.  This scaling is consistent with equating the growth timescale of the linear mode ($\sim \beta^{-1}$) 
with the libration period of nonlinearly trapped orbits ($\propto \vert \delta \Sigma\vert^{-1/2}$).
The possibility that trapping is somehow responsible for saturation has also been mentioned in passing by several other authors (e.g. \citealt{kalnajs1977dynamics, donner1994structure}), but the precise mechanism behind the saturation, and a quantitative estimate of the resulting amplitude, has been lacking.

In this paper we argue quite generally that saturation of spiral instabilities may well involve nonlinear trapping at corotation, 
which crucially \textit{flattens} the angular momentum DF in the vicinity of the resonance, erasing the sharp features that allowed for instability. We derive an analytical formula for the saturated amplitude of a spiral mode whose growth rate is very small --- i.e. $\beta/\Omega_\mathrm{p} \ll 1$, where $\Omega_\mathrm{p}$ is the pattern speed --- and recover the scaling $\delta \Sigma \propto \beta^2$. While \cite{sellwood2022spiral}'s spirals do not satisfy the condition $\beta/\Omega_\mathrm{p} \ll 1$ particularly well (in a sense we will define precisely below),
the physics behind their saturation is essentially the same as that captured by our calculation, so their spirals obey the same scaling albeit with a modified prefactor.

Inspiration for our analytic calculation was primarily drawn from the classic plasma papers by \cite{ONeil1965-uy} and \cite{Dewar1973-rf}.  In the galactic context, the only comparable calculation we know of is by \citet{Morozov1980-pm} (later reproduced as \S2.2 of the book by \citealt{Fridman1984}), 
who studied the saturation of amplified spiral perturbations in a \textit{stable} galactic disk.

This paper is structured as follows.  In \S\ref{sec:calculation} we explain the physics behind spiral mode amplification and saturation, and thereby calculate the resulting saturation amplitude and estimate the corresponding relative surface density perturbation
(some mathematical details are relegated to the Appendix).  In \S\ref{sec:Discussion} we compare our results with those of \cite{sellwood2022spiral} and with the classic problem of bar-halo friction, and discuss briefly the astrophysical implications of our work.
We summarize in \S\ref{sec:Summary}.


\section{Calculating the saturation amplitude}
\label{sec:calculation}

\begin{figure}
    \centering
    \includegraphics[width=0.34\textwidth]{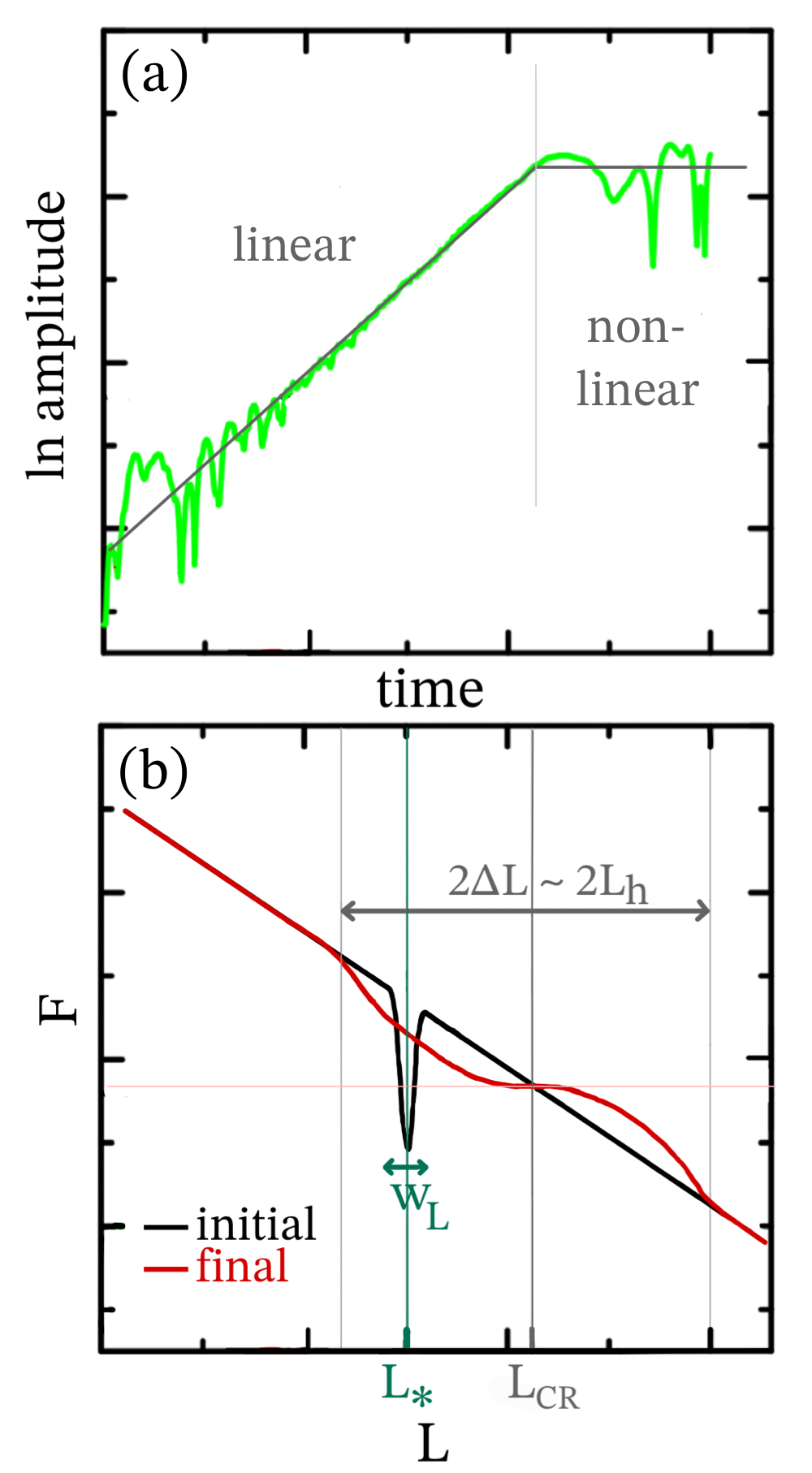}
    \caption{(a) Growth and saturation of a spiral mode in an $N$-body simulation of a razor thin stellar disk (adapted from \citealt{sellwood2022spiral}, green line). The gray line illustrates the simplification used in this paper: we separate the spiral's life into a linear phase of perfectly exponential growth, and a nonlinearly saturated phase of constant amplitude. (b) Illustration of the change to the axisymmetric DF $F(L, t)$ in the vicinity of a corotation resonance.
    The initial instability is driven by some sharp feature in the DF (here a \textit{groove}, centered on $L=L_*$ with width $w_\mathrm{L}$, see \citealt{sellwood2022spiral}).
    In the saturated state the DF is flattened around the resonance $L = L_\mathrm{CR}$.}
    \label{fig:Amplitude}
\end{figure}

To establish the basic idea behind our calculation, we first present 
Figure \ref{fig:Amplitude}. The green line in panel (a), which is adapted from Figure 6 of \cite{sellwood2022spiral}, 
shows the logarithmic amplitude of a spiral density wave from their $N$-body simulation of an initially grooved, razor thin disk.
The gray line illustrates schematically the simplification we will use here:
we pretend that the evolution
consists of a linear phase in which the spiral amplitude grows exponentially, followed immediately by a
nonlinear phase in which the saturated spiral has constant amplitude.
In panel (b) we sketch the distribution function (DF) of angular momenta in the disk, $F(L)$.
The initial unstable distribution is shown with a black line.
Saturation occurs because nonlinear trapping tends to flatten the DF in the vicinity of corotation (red line), erasing the feature that drove the instability. By calculating the angular momentum content of the spiral in the linear (\S\ref{sec:linear}) and nonlinear (\S\ref{sec:nonlinear}) regimes, and equating the two, we will be able to read off an approximate expression for the saturation amplitude (\S\ref{sec:saturation}).

It is worth noting that unstable modes in many different physical systems exhibit very similar behavior to that shown in Figure \ref{fig:Amplitude}.
This is true not only for other unstable stellar disks
(\citealt{donner1994structure}) but also for stellar clusters prone to radial orbit instabilities
\citep{palmer1994stability,petersen2023predicting}, for various unstable plasmas \citep{fried1971nonlinear,Berk1995-hk,lesur2020nonlinear,bierwage2021effect},
 and in grooved/ridged Hamiltonian Mean Field (HMF) models (\citealt{campa2017quasilinear}; Modak \& Hamilton, in prep).  Thus, our calculation can, with minor adjustments, be used to predict saturation 
amplitudes in a whole class of unstable kinetic systems.

\subsection{Model}

We consider a two-dimensional, razor-thin stellar disk, which we embed in a rigid dark matter halo so as to render the disk only weakly unstable at $t=0$.
Let the gravitational potential of the whole system be $\Phi$, which we decompose into a dominant axisymmetric background part $\Phi_0$, and a non-axisymmetric perturbation $\delta \Phi$ which is generated by the non-axisymmetric part of the stellar distribution.

In the disk plane $(\varphi, R)$ we can describe stellar orbits using angle-action coordinates \citep{Binney2008-ou}:
\begin{equation}
    \btheta = (\theta_\varphi, \theta_R), \,\,\,\,\,\,\,\,\,\,\, \bJ = (J_\varphi, J_R),
\end{equation}
where $J_R$ is the radial action and $J_\varphi = R^2 \md \varphi/\md t \equiv L$ is the angular momentum. A star orbiting in this disk has Hamiltonian
\begin{equation}
    H(\btheta,\bJ,t) = H_0(\bJ) + \delta \Phi(\btheta,\bJ,t),
    \label{eqn:H}
    \end{equation}
   where $ H_0 = \mathbf{v}^2/{2} + \Phi_0(\mathbf{x})$
and $(\mathbf{x},\mathbf{v})$ are the star's position and velocity. The orbital frequencies are
\begin{equation}
    \boldsymbol{\Omega}(\bJ) \equiv \frac{\partial H_0}{\partial \bJ} = (\Omega_\varphi, \Omega_R).
\end{equation}

Note that stars on nearly-circular orbits --- which is approximately true of most stars in a rotationally-supported, axisymmetric disk --- have $J_R \approx 0$, $\theta_\varphi \approx \varphi$ and $\bOm \approx \bOm(L)$.
In particular, when making various estimates in this paper we will consider the specific case of a Mestel disk, which is also the model used by \cite{sellwood2022spiral}.
In this model, stars on circular orbits have a flat rotation curve:
\begin{equation}
    v_\varphi(J_R = 0, L)  = R\,  \Omega_\varphi(J_R=0, L) = V_0 = \mathrm{const}.
\end{equation}
It follows that for circular orbits, $L = RV_0$ and so 
\begin{equation}
    \Omega_\varphi(0,L) = \frac{V_0}{R} = \frac{V_0^2}{L}.
    \label{eqn:Omega_Mestel}
\end{equation}

We also define the distribution function (DF) of stars in the disk, $f$, such that $\int \md^2 \mathbf{v} \,f = \Sigma(\mathbf{x}, t)$ is the stellar surface density.
We further decompose this into an axisymmetric part and a non-axymmetric part:
\begin{equation}
    f(\btheta,\bJ,t) = F(\bJ, t) + \delta f(\btheta,\bJ,t), \end{equation}
where
\begin{equation}
    F(\bJ, t) \equiv \frac{1}{(2\pi)^{2}} \int \md \bm{\theta}\, f(\btheta, \bJ, t),
\end{equation}
and the corresponding density perturbation is $\delta \Sigma(\mathbf{x}, t) = \int \md^2\mathbf{v} \,\delta f$.


\subsection{Linear regime}
\label{sec:linear}

Now, at $t=0$ we suppose the axisymmetric part of the DF is $F(\bJ, 0) = f_0(\bJ)$, and that this DF admits an unstable, $m$-armed, global spiral Landau mode with growth rate $\beta$,
which rotates rigidly in the $\varphi$ direction with pattern speed $\pattern = \omega/m$ (so the complex mode frequency is $\omega + i \beta$).
We will not attempt to calculate $\beta$ or $\omega$ self-consistently in this paper,
but this can be done using linear response theory (e.g. \citealt{evans1998stability,jalali2005unstable,de2016spiral,De_Rijcke2019-uo,petersen2023predicting}).
This spiral mode will exponentiate out of the initial non-axisymmetric fluctuation\footnote{Or more precisely, the projection of this fluctuation onto the density field of the mode in question.} $\delta f(t=0)$, which can either be imposed by hand or can be due to some initial (e.g. finite-$N$) noise. After an early transient phase the exponentially growing mode will dominate the potential, so for $t\gtrsim \beta^{-1}$ we can write 
\begin{align}
    \delta \Phi(\varphi, R, t) &=  A(R) \exp(\beta t) \cos\{ m[\varphi - \pattern t + g(R)]\},
    \label{eqn:exp_growing}
\end{align}
for some shape function $m g(R)$. 
Expanding this as a Fourier series in angles, $\delta \Phi = \sum_{\mathbf{n}}\exp(i\mathbf{n}\cdot\bm{\theta})\delta\Phi_{\mathbf{n}}(\bJ, t)$ for $\mathbf{n} \in \mathbb{Z}^2$, we can identify\footnote{Here, 
\begin{equation}
    \psi_{\mathbf{n}}(\bJ) \equiv \pi \int \frac{\md \theta_R}{2\pi} \exp(-in_R\theta_R) A(R) \exp\{ in_\varphi [\lambda(R) + g(R)]\},
\end{equation}
where $\lambda(R) \equiv \varphi - \theta_\varphi$, and we used the fact that $R = R(\theta_R, \bJ)$ is independent of $\theta_\varphi$.}
\begin{align}
    \delta \Phi_\mathbf{n}(\bJ, t) = \begin{cases}
        \psi_{\mathbf{n}}(\bJ) \exp(\beta t) \exp(-i n_\varphi \pattern t), \,\,\,\,\, &n_\varphi = \pm m,
         \\
     0, \,\,\,\,\, &\mathrm{otherwise}.
    \end{cases}
    \label{eqn:deltaPhinJt}
\end{align}

Now, the rate of angular momentum transfer to a population of stars
by a potential perturbation $\delta \Phi$ is
\begin{align}
    \frac{\md \mathcal{L}}{\md {t}} &= - \int \md \btheta \, \md \bJ \, \frac{\partial \delta \Phi(\btheta, \bJ, t)}{\partial \theta_\varphi}  f(\btheta,\bJ,t) \nn
    \\
    &= i(2\pi)^2\sum_{\bm{n}}n_\varphi \int \md \bJ \, \delta f_{\mathbf{n}}(\bJ,t) \, \delta \Phi^*_{\mathbf{n}}(\bJ, t).
    \label{eqn:total_torque}
\end{align}
Assuming $\delta \Phi$ is weak and $F = f_0$ is time-independent, we can solve 
the linearized collisionless Boltzmann equation $\partial \delta f_\mathbf{n}/\partial t + i\mathbf{n}\cdot\bOm \delta f_\mathbf{n} - i (\mathbf{n} \cdot \partial  f_0/\partial \bJ) \delta \Phi_\mathbf{n} = 0$ for $\delta f_\mathbf{n}$ in terms of $\delta \Phi_{\mathbf{n}}$. Plugging the result into \eqref{eqn:total_torque}, using the form \eqref{eqn:deltaPhinJt} for $\delta \Phi_{\mathbf{n}}$, 
and again ignoring terms that are exponentially subdominant for $t\gtrsim \beta^{-1}$, 
we arrive at (see \citealt{,dekker1976spiral} for details):
\begin{equation}
    \frac{\md \mathcal{L}}{\md {t}} =  2m \exp(2\beta t)  \, \mathrm{Im} \left[ M_m(\omega + i \beta) \right],
    \label{eqn:ang_mom_transfer}
\end{equation}
where $\omega = m\pattern$ and 
\begin{align}
    M_m(z) \equiv  (2\pi)^2 & \sum_{n}
    \int  \md L \, \md J_R 
    \nn
    \\
& \times    \frac{m \, \partial f_0/\partial L + n\, \partial f_0/\partial J_R}{z - (m\Omega_\varphi + n \Omega_R)}
    \vert  \psi_{mn}(\bJ)\vert^2
\end{align}
and we used the shorthand $n_R = n$.
We recognise $M_m(z)$ as closely related to the \textit{response matrix} (e.g. \citealt{kalnajs1977dynamics}) which encodes the self-consistent response of the system to $m$-armed perturbations of frequency $z$ (with Im $z>0$). In particular, 
if \eqref{eqn:exp_growing} is a true Landau mode of the system then $\mathrm{Im}\,[M_m(\omega + i \beta)] = 0$.
From \eqref{eqn:ang_mom_transfer},
this guarantees total angular momentum conservation $\md \mathcal{L}/\md t = 0$ in the linear approximation.

To gain further insight let us expand the right hand side of \eqref{eqn:ang_mom_transfer} for small $\beta/\omega \ll 1$. This gives
\begin{align}
    \frac{\md \mathcal{L}}{\md {t}} = \dot{\mathcal{L}}_\mathrm{res} + \dot{\mathcal{L}}_\mathrm{non-res} + \mathcal{O}[(\beta/\omega)^2],
        \label{eqn:ang_mom_transfer_wdl}
    \end{align}
    where
    \begin{align}
    &\dot{\mathcal{L}}_\mathrm{res} \equiv  2m \exp(2\beta t)  \, \mathrm{Im}\,  M_m(\omega), 
    \\
    &\dot{\mathcal{L}}_\mathrm{non-res} \equiv 2m\beta  \exp(2\beta t)  \, \mathrm{Re}\,  M'_m(\omega),
\end{align}
and $M_m'(\omega) \equiv [dM_m/dz ]_{z=\omega}.$
The terms $\dot{\mathcal{L}}_\mathrm{res}$ and $\dot{\mathcal{L}}_\mathrm{res}$ correspond to
angular momentum transferred to the resonant and non-resonant stars respectively. For instance, writing $\mathrm{Im}\,   M_m(\omega ) = \mathrm{Im} \lim_{\beta \to 0^+} M_m(\omega + i \beta)$ and using Plemelj's formula it is easily to show that
\begin{align}
    \dot{\mathcal{L}}_\mathrm{res} = -2 (2\pi)^2 m & \exp(2\beta t) \sum_{n}\int \md L \, \md J_R 
    \left( m\frac{\partial f_0}{\partial L} + n\frac{\partial f_0}{\partial J_R}\right) 
 \nn   \\
 \times & \vert \psi_{mn} (\bJ)\vert^2
    \pi \delta(m\pattern - m\Omega_\varphi + n\Omega_R).
    \label{eqn:Ldotres}
\end{align}
This expression involves only those stars whose orbital frequencies are exactly resonant with the spiral mode.
On the contrary, $\dot{\mathcal{L}}_\mathrm{non-res}$ does not involve the resonant stars (its $\bJ$ integral involves a principal value at each resonance location).

Since the total angular momentum is conserved, we see from \eqref{eqn:ang_mom_transfer_wdl} that to first order in $\beta/\omega$ the 
 angular momentum gained/lost by the resonant stars is lost/gained by the non-resonant stars,
$\dot{\mathcal{L}}_\mathrm{non-res} \approx - \dot{\mathcal{L}}_\mathrm{res}$.
Physically, the non-resonant stars, which make up the bulk of the system, oscillate in such a way as to sustain the spiral mode \eqref{eqn:exp_growing},
while the resonant stars, which make up a small subset of the system, feed angular momentum to the mode and cause it to grow \citep{Nelson1999-in,vandervoort2003stationary}.
We can therefore identify $\mathcal{L}_\mathrm{mode}(t) \equiv \int_0^t \md t' \dot{\mathcal{L}}_\mathrm{non-res}(t') \approx -  \int_0^t \md t' \dot{\mathcal{L}}_\mathrm{res}(t')$ as the `angular momentum of the spiral mode'.\footnote{In truth, whether we give 
$\mathcal{L}_\mathrm{mode}$ the name `spiral angular momentum' is only a matter of semantics.
Really, `$\mathcal{L}_\mathrm{mode}$' is just a shorthand for `the total excess angular momentum carried by stars whose orbital frequencies are not within a range $\vert \Delta \bOm \vert \sim \beta$ of the corotation resonance' (see the last paragraph of \S\ref{sec:linear}).
}
Integrating \eqref{eqn:Ldotres} forward in time we get for $t\gg \beta^{-1}$:
\begin{align}
    {\mathcal{L}}_\mathrm{mode} \approx  \frac{4\pi^3 m}{\beta} \exp(2\beta t) & \sum_{n}\int \md L \, \md J_R 
    \left( m\frac{\partial f_0}{\partial L} + n\frac{\partial f_0}{\partial J_R}\right) 
 \nn   \\
 \times & \vert \psi_{mn} (\bJ)\vert^2
    \delta(m\pattern - m\Omega_\varphi + n\Omega_R).
    \label{eqn:Energy_Wave}
\end{align}
It is found empirically that the dominant resonance responsible for angular momentum transfer amplifying spiral modes (at least for groove-induced modes, see \citealt{sellwood1991spiral,palmer1994stability} and the discussion in \S\ref{sec:SC22}) is the corotation resonance $\Omega_\varphi = \pattern$.  Therefore we will keep only the contributions from $n=0$ in \eqref{eqn:Energy_Wave}.
Performing the resulting integral
over $L$ we find 
\begin{align}
    \mathcal{L}_\mathrm{mode} & \approx \frac{4\pi^3 m}{\beta} \int \md J_R \left[ \frac{\vert \Phi_m \vert^2}{\vert \partial \Omega_\varphi/\partial L \vert } \pder{f_0}{L}\right]_{L =  L_\mathrm{CR}(J_R)},
    \label{eqn:linear_AM}
    \end{align}
        where
    $L_\mathrm{CR}(J_R)$ is the angular momentum value corresponding to the  corotation resonance at a given $J_R$:
        \begin{equation}
        \Omega_\varphi(J_R, L_\mathrm{CR}(J_R)) = \Omega_\mathrm{p},
        \label{eqn:Corotation_Resoannce}
    \end{equation}
    and
    \begin{equation}
        \vert \Phi_m \vert \equiv \vert \psi_{m0}(\bJ) \vert \exp(\beta t)
    \end{equation}
is the current magnitude of the $m$-th Fourier component of the potential fluctuation \eqref{eqn:deltaPhinJt}.

The expressions \eqref{eqn:Energy_Wave}-\eqref{eqn:linear_AM} are of zeroth order in the small parameter $\beta/\omega$.
More generally, if we kept higher order terms in \eqref{eqn:ang_mom_transfer_wdl} we would find that the finite value of $\beta$ broadens the $\delta$-function in \eqref{eqn:Energy_Wave} so that resonant stars occupy a width in orbital frequency space of $\vert \Delta \bOm \vert \sim \beta$. 
Converting this to angular momentum space for circular orbits in the Mestel disk gives a corotation resonance width according to linear theory of
\begin{align}
    \vert \Delta L\vert  &\sim \bigg\vert \frac{\partial \Omega_\varphi}{\partial L} \bigg\vert^{-1}_\mathrm{CR} \vert  \Delta \bOm \vert \sim
    \frac{L_\mathrm{CR}^2}{V_0^2} \beta = \frac{\beta}{\pattern} L_\mathrm{CR}, \label{eqn:resonance_width_linear}
\end{align}
where $\Omega_\varphi(L_\mathrm{CR}, 0) = \pattern$.
This will be important when interpreting the results of \cite{sellwood2022spiral} in \S\ref{sec:SC22}.

    \subsection{Nonlinear regime}
\label{sec:nonlinear}

The above results rely on linear theory, i.e. they assume that stellar orbits are only slightly --- and sinusoidally --- nudged off their unperturbed trajectories by the spiral potential $\delta \Phi$.  
This implies that any changes to the angle-averaged DF $F$ are of second order in the spiral amplitude and so can be ignored to the accuracy we require here.
However, sufficiently close to resonance and on sufficiently long timescales, stars will in fact be trapped onto qualitatively different orbits, such that when viewed in the rotating frame of the spiral they librate around a fixed angle.
Mathematically, for the corotation resonance this motion can be described using a pendulum equation in the variables $(\theta_\varphi - \pattern t, L)$ --- see the Appendix for details. 
In particular, the
period of the nonlinear librations is $t_\mathrm{lib} = 2\pi/\omega_\mathrm{lib}$, where $\omega_\mathrm{lib}$ is given in equation \eqref{eqn:t_libration}. The maximum extent in $L$ of the `island' of  trapped orbits at corotation is given by the island half-width $L_\mathrm{h}$ --- see equation \eqref{eqn:half_width}. 

Suppose for a moment that the spiral amplitude is held constant. Then in and around the librating island the full DF $f(\btheta, \bJ, t)$ gets progressively more sheared (phase-mixed) along the pendulum Hamiltonian contours. {As a result, once $t\gg t_\mathrm{lib}$ the DF is a function only of the pendulum Hamiltonian itself (so is uniform along any given Hamiltonian contour), and is independent of oscillation phase (for trapped orbits, this means independent of libration angle)} --- see e.g. \S3.3.3 of \citealt{Hamilton2023}.
Figure \ref{fig:Amplitude}b shows the corresponding effect that this phase mixing has on the 
angle-averaged DF of resonant stars $F(L,t)$.  For $t\to\infty$ there is a significant change to $F$ in the libration region $\vert L - L_\mathrm{CR} \vert  \lesssim L_\mathrm{h}$ compared to $t=0$, but little change outside of this region.
Most importantly, $F$ has \textit{flattened} in the vicinity of the 
resonance. 

The above assumes the spiral amplitude is fixed, whereas in truth it is growing, as is the width of the trapped island: $L_\mathrm{h} \propto \vert \Phi_m \vert^{1/2} \propto \exp(\beta t/2)$. However, the phase mixing occurs on the timescale $\sim \omega_\mathrm{lib}^{-1}$, and as we will see, around the time that the spiral saturates this is actually $\lesssim \beta^{-1}$, so the spiral amplitude may be considered quasi-static\footnote{One may worry that for stars near the separatrix between librating and circulating orbits, $\omega_\mathrm{lib}^{-1}$ formally diverges.
However this is a weak (logarithmic) divergence, and applies to such a small fraction of stars that we do not expect it to  affect our simple estimates.}.
Then, the flattening of the axisymmetric part of the DF in the vicinity of the resonance quenches the gradients that were responsible for the mode amplification (e.g. by filling in the groove, see Figure \ref{fig:Amplitude}b), and so the perturbations cease to grow and the system reaches a stationary state.\footnote{More precisely, 
one can show that the resulting phase-mixed DF $f$ (which contains both axisymmetric and non-axisymmetric parts) is an approximate soliton; that is, a self-consistent, nonlinear, rigidly-rotating solution to the combined Vlasov-Poisson system of equations, much like a stellar bar \citep{vandervoort2003stationary}. }
Just as in \S\ref{sec:linear}, we can calculate the angular momentum excess contained in the nonresonant stars in this saturated state (i.e. the angular momentum of the spiral mode) by setting it equal to minus the change in the angular momentum of the resonant stars compared to $t=0$.
In the Appendix we show that this is approximately
\begin{align}
    \mathcal{L}_\mathrm{mode} \approx  \frac{512 \pi \sqrt{2m}}{9} \int \md J_{R} \left[ \frac{\vert \Phi_m \vert^{3/2}}{\vert \partial \Omega_\varphi / \partial L \vert^{3/2}}  \, \pder{f_0}{L }\right]_{L = L_\mathrm{CR}(J_R)}.
\label{eqn:Delta_L_final}
\end{align}

\subsection{Saturation amplitude}
\label{sec:saturation}

Following Figure \ref{fig:Amplitude}a, we now equate the angular momentum in the spiral immediately before saturation \eqref{eqn:linear_AM} with that after saturation \eqref{eqn:Delta_L_final}.
This requirement leads to an equation of the form $\int \md J_R \, Q(\Phi_m, J_R) = 0$.  However, neither linear or nonlinear perturbations at corotation are able to change a star's radial action, so we must have $Q=0$ at \textit{every} $J_R$.
This allows us to read off an expression for the final saturation amplitude of the spiral potential at corotation:
\begin{equation}
    \vert \Phi_m \vert =  \left( \frac{128}{9\pi^2} \right)^2 \frac{2}{m}\, \frac{\beta^2}{\vert \partial \Omega_\varphi/\partial L \vert_{L_\mathrm{CR}}}.
\label{eqn:Saturation_Amplitude}
\end{equation}
This result can be made more physically enlightening if we use \eqref{eqn:F_and_G} and \eqref{eqn:t_libration} to express it as 
\begin{equation}
   \omega_\mathrm{lib} =  \frac{256}{9\pi^2} \, m^{1/2} \beta \approx 2.88\, m^{1/2} \beta.
   \label{eqn:Libration_Saturation}
\end{equation}
In other words, saturation occurs when the nonlinear libration frequency $\omega_\mathrm{lib}$ is a few times larger than the 
linear growth rate $\beta$. 
Using equations \eqref{eqn:t_libration}-\eqref{eqn:half_width} we can estimate the final island width to be 
\begin{equation}
     L_\mathrm{h} \sim \bigg\vert \frac{\partial \Omega_\varphi}{\partial L} \bigg\vert^{-1}_\mathrm{CR}
    \omega_\mathrm{lib}  \sim \frac{L_\mathrm{CR}}{\pattern}\beta \sim \frac{L_\mathrm{CR}^2}{V_0^2}\beta,
    \label{eqn:Lh_sat}
\end{equation}
which is the same as the linear width $\Delta L$ (equation \eqref{eqn:resonance_width_linear}).
Thus, the librating island of nonlinearly trapped orbits simply grows until it engulfs the original resonance width implied by time-dependent linear
theory. 
Spiral instabilities with higher growth rates thus saturate at larger amplitudes.

 Note that we did not have to consider explicitly any `groove' or other feature in the DF in order to arrive at equation \eqref{eqn:Saturation_Amplitude}. The only requirement was that \textit{some} feature in the DF causes it to be unstable, and that this instability can be quenched by flattening the DF in the vicinity of the corotation resonance (Figure \ref{fig:Amplitude}).
 Note also that for one-armed spirals, equation \eqref{eqn:Libration_Saturation} is identical to equation (12) of \cite{Dewar1973-rf}, who studied the saturation of \textit{linear} momentum transfer between electrons and a single longitudinal plasma wave.

Let us evaluate equation \eqref{eqn:Saturation_Amplitude} for the Mestel disk.
Using \eqref{eqn:Omega_Mestel} we find
\begin{equation}
    \vert \Phi_m \vert =  \left( \frac{128}{9\pi^2} \right)^2 \frac{2}{m}\, \frac{V_0^2 \beta^2}{\Omega_\mathrm{p}^2}.
\label{eqn:Saturation_Amplitude_Mestel}
\end{equation}
\begin{figure}
    \centering
    \includegraphics[width=0.45\textwidth]{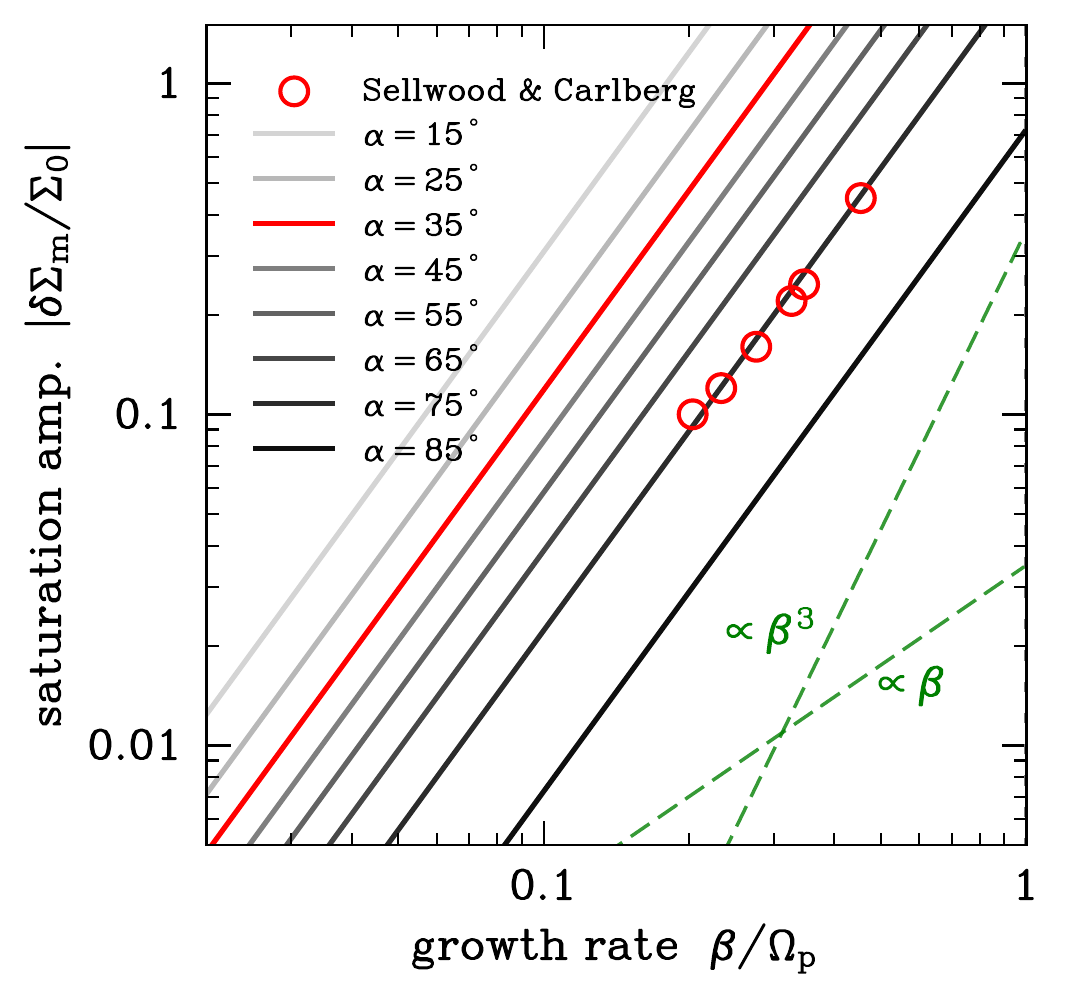}
    \caption{Saturation amplitude of spiral instabilities at corotation as a function of dimensionless growth rate.
    The theoretical result \eqref{eqn:Rel_Surf} is shown for $\xi=1/2$ and different values of pitch angle $\alpha$ (solid lines).
    The results of \citet{sellwood2022spiral}'s simulations of $m=2$ spirals are shown with red circles (c.f. their Figure 11).}
    \label{fig:SC22}
\end{figure}
We can further relate this to a relative surface density perturbation at corotation, $\vert \delta \Sigma_m(R_\mathrm{CR})/\Sigma_0\vert$.
Using equation \eqref{eqn:Saturation_Amplitude_Mestel} we can easily estimate this to be of order 
$\sim 
  \vert \Phi_m \vert /V_0^2  
   \sim  4 m^{-1}  (\beta/\Omega_\mathrm{p})^2$.  
   To make this estimate more precise, we must relate the potential of the spiral to the corresponding 
   surface density perturbations.  This can be done in the WKB approximation provided the radial wavelength $\lambda$ of the spiral is not too large; in this case $\vert \Phi_m \vert \approx  G \lambda \vert \,\delta \Sigma_m\vert$ (\citealt{Binney2008-ou}, equation (6.30)).
   We now divide this by the surface density of the Mestel disk $\Sigma_0 = \xi V_0^2/(2\pi GR)$, where
   $\xi$ is the disk mass fraction (so $\xi = 1/2$ for a half-mass disk).
   Evaluating at the corotation radius, 
we find
\begin{align}
    \bigg \vert \frac{\delta \Sigma_m}{\Sigma_0} \bigg \vert &= \frac{32768}{81\pi^4} \, \xi^{-1} \cot \alpha \left( \frac{\beta}{\Omega_\mathrm{p}} \right)^2
    \label{eqn:Rel_Surf}
    \\
    &\approx 0.5 \times \left(\frac{\xi}{0.5}\right)^{-1} \left(\frac{\cot \alpha}{1.5}\right) \left( \frac{\beta/\Omega_\mathrm{p}}{0.2} \right)^{2},
        \label{eqn:Rel_Surf_numerical}
\end{align}
where $\alpha \equiv \arctan[\lambda m/(2\pi R_\mathrm{CR})]$ is the pitch angle at corotation.  Note that when written in this form, \eqref{eqn:Rel_Surf} does not depend explicitly on $m$.



\section{Discussion}
\label{sec:Discussion}

\subsection{Comparison with Sellwood \& Carlberg (2022)}
\label{sec:SC22}

\cite{sellwood2022spiral} considered spiral saturation in $N$-body simulations of initially axisymmetric half-mass Mestel disks \citep{Binney2008-ou}. They generated the instabilities by carving a groove in the DF of these disks centered on a particular location in angular momentum space at $t=0$, namely $L_* \approx 6.5R_0V_0$ (where $R_0$ is their length unit) --- see Figure \ref{fig:Amplitude}b for a schematic illustration.
By varying the width of the groove or the level of random motion in the disk, they were able to generate spirals with very similar pattern speeds $\Omega_\mathrm{p}$ but different growth rates $\beta$.  Their Figure 11 shows the saturation amplitude of the resulting spiral instabilities as a function of $\beta/\Omega_\mathrm{p}$.

To compare our results with those of \cite{sellwood2022spiral}, 
in Figure \ref{fig:SC22} we plot $\vert \delta \Sigma_m / \Sigma_0\vert $ as a function of 
the dimensionless growth rate $\beta/\Omega_\mathrm{p}$, following equation \eqref{eqn:Rel_Surf}, for different pitch angles $\alpha$. With red markers we show the 6 data points for the $m=2$ spirals extracted by \cite{sellwood2022spiral} in their Figure 11. 
We see that the theory is an excellent fit to the data for $\alpha \approx 75^\circ$.
However, inspection of \cite{sellwood2022spiral}'s figures suggests their $m=2$ spirals have pitch angles closer to $\alpha \sim 35^\circ$, in which case the prediction \eqref{eqn:Rel_Surf} is consistently too large by a factor of a few, although the scaling $\vert \delta \Sigma_m / \Sigma_0\vert \propto (\beta/\Omega_\mathrm{p})^2$ remains accurate.


Why might our the theory be systematically overestimating \cite{sellwood2022spiral}'s spiral amplitudes?
The main reason is 
that \eqref{eqn:Rel_Surf} was derived in the limit of very small $\beta/\pattern$,
and this is not a good approximation
for \cite{sellwood2022spiral}'s spirals.
One way to see this clearly is to note that
in this limit, all nonzero contributions
to the angular momentum transfer between resonant and non-resonant stars in the linear regime
occur at exact resonant locations in action space (see equation \eqref{eqn:Energy_Wave}).
This cannot be true for \cite{sellwood2022spiral}'s spirals, since the groove  (centered on $L_* = 6.5R_0V_0$ and with a width of $w_L \sim 0.2R_0V_0$) does not coincide with the corotation resonance ($L_\mathrm{CR} \approx 7.1 R_0V_0$ for $m=2$)\footnote{{The fact that $L_\mathrm{CR}>L_*$ is a generic feature of low-$m$ groove-induced spiral modes \citep{sellwood1991spiral}.}} or any other major resonance of the half-mass Mestel disk
(see the top panel of \citealt{sellwood2022spiral}'s Figure 4).
In other words,  if we were to calculate angular momentum transfer using equation \eqref{eqn:Energy_Wave} using their grooved DF, the grooved region itself would contribute almost nothing, which is clearly absurd. Instead, a more accurate calculation of the mode's angular momentum content in {the} linear regime should involve an integral over each broadened resonance, with width (equation \eqref{eqn:resonance_width_linear}):
\begin{align}
    \vert \Delta L\vert  &\sim  7.1\frac{\beta}{\pattern} \times  R_0 V_0.
\end{align}
This broadened function, when centered on the corotation resonance, engulfs the groove region
for all six of \citet{sellwood2022spiral}'s $\beta/\pattern$ values,
so the mode \textit{can} feed off the sharp gradients in the DF created by the groove, 
as illustrated in Figure \ref{fig:Amplitude}b.

Despite this subtlety, the physics of mode saturation is essentially the same as in our simplified calculation: nonlinear trapping at corotation and subsequent phase mixing flattens the DF, filling in the grooved region and hence quenching the instability. This explains why \cite{sellwood2022spiral} still find the scaling $\vert \delta \Sigma_m /\Sigma_0 \vert \propto \beta^2$ predicted by \eqref{eqn:Saturation_Amplitude}.
The appropriate prefactor in their case is smaller, likely because to quench the instability it {is} not necessary for the DF to be completely phase mixed in the resonant region; it is sufficient merely to fill in  a significant portion of the groove.

\subsection{Comparison with bar-halo friction}

It is tempting to draw an analogy between our calculation and the classic problem of bar-halo friction in galaxies \citep{Tremaine1984-wt,chiba2022oscillating,chiba2023dynamical,Hamilton2023}. These problems are indeed {closely related:} in both cases, angular momentum is {transferred} to/from a global, rigidly rotating `mode' (spiral or bar), and the equations governing this transfer in the linear regime (\S\ref{sec:linear}) look very similar.  
Moreover, in both cases the angular momentum transfer can be quenched due to the trapping of orbits at (e.g.) corotation.
However, the precise physics involved is subtly different.
{To illustrate this, let us consider the simple case of a constant-amplitude bar whose pattern speed $\pattern$ is decreasing monotonically.}
{The decrease in $\pattern$ causes e.g. the corotation resonance location to sweep to larger radii (larger angular momenta $L_\mathrm{CR}$).
In particular, if the resonance location moves by a resonance width $\sim L_\mathrm{h}$ on a timescale short compared to the libration time of trapped orbits $t_\mathrm{lib}$ (the so-called `fast regime', see \citealt{Tremaine1984-wt}), then the bar will always be encountering fresh, untrapped material with which to resonate. Mathematically, in this regime nonlinear trapping can be ignored and linear theory will continue to be valid (see \citealt{chiba2023dynamical} for a general treatment).}
On the other hand, in the case of spiral Landau modes, the phase space location of the corotation resonance is fixed, and only its width grows with time until eventually it is comparable to the width implied by the finite $\beta$ in linear theory (compare equations \eqref{eqn:resonance_width_linear}-{\eqref{eqn:Lh_sat}). Thus in the absence of diffusive processes, secular evolution, etc. (see \S\ref{sec:Further_Discussion}), amplifying spirals never encounter `fresh material' at all. 
They simply grow until the trapping region has exhausted the initial population of resonant orbits (within $\vert \Delta \bOm \vert  \sim \beta^{-1}$ of the resonance) off of which the spiral was feeding.

\subsection{Astrophysical implications}
\label{sec:Further_Discussion}

Saturation amplitudes of spirals are of key astrophysical importance because they may allow one to distinguish between 
different theories of spiral structure \citep{d2013self,sellwood2021spiral}.
More generally, spiral strength is a key parameter in determining the secular evolution of galactic disks, their associated gas flows, star formation rates, etc. \citep{roberts1969large,Binney1988-zy,Sellwood2002-lv,slyz2003exploring,kim2014gaseous}.

Taken at face value, the result \eqref{eqn:Rel_Surf} suggests that spiral structure driven by instabilities will only ever reach a relative amplitude of several tens of percent compared to the axisymmetric component of a galaxy.  This is because (i) most spirals have $\cot \alpha \lesssim $ a few \citep{lingard2021galaxy}, and (ii) it is difficult to imagine any
sustainable cycle of instabilities whose individual mode strengths amplify faster than $\sim \pattern$, so typically $(\beta/\pattern)^2 \ll 1$.
This result is consistent with observations of isolated galaxies; overdensities of $100\%$ or greater are mostly confined to galaxies undergoing strong tidal interactions \citep{elmegreen1989spiral, rix1993tracing, kendall2011spiral, kendall2015spiral}.

There may, however, exist a slight tension between theory and observation given that equation \eqref{eqn:Rel_Surf} predicts that more tightly-wrapped spirals will have larger amplitudes, whereas \cite{diaz2019shapes} and \cite{yu2020statistical}
found (very weak) correlations in the opposite sense in S$^4$G and SDSS data respectively.
On the other hand, $\beta$, $\pattern$ and $\alpha$ are emphatically \textit{not} independent variables, and since self-gravity is weaker for tighter spirals, a decrease in $\alpha$ will correspond to a decrease in $\beta/\pattern$.

It is enlightening to estimate the timescale $t_\mathrm{sat}$ over which saturation of realistic spirals might be achieved. To do this, first recall that saturation occurs when $\omega_\mathrm{lib}(t_\mathrm{sat}) \sim \beta$.  Second, a basic scaling of equation \eqref{eqn:t_libration} tells us that at corotation, $\omega_\mathrm{lib}(t) \sim  \epsilon(t)^{1/2}\pattern$, where $\epsilon(t) \equiv \vert  \Phi_m(t) /\Phi_0\vert$ is the dimensionless strength of the perturbation.  Third, in the linear regime we have  $\epsilon(t) =\epsilon(0)\exp(\beta t)$.
Putting these three ingredients together, we get a saturation time
\begin{equation}
     t_\mathrm{sat} \sim  \frac{2}{\beta} \left[ \ln \left( \frac{\beta}{\pattern}\right)  + \frac{1}{2} \ln \left( \frac{1}{\epsilon(0)}\right)\right].
\end{equation}
In particular, if the initial fluctuation level is set by Poisson noise alone then $\epsilon(0) \sim N^{-1/2}$, and we have (very roughly):
\begin{align}
     t_\mathrm{sat} 
     &\sim  \frac{1}{2\beta}  \ln N \\
     & \sim 2.5\, \mathrm{Gyr} \times \left( \frac{T_\mathrm{CR}}{250 \, \mathrm{Myr}} \right)
    \left( \frac{\beta/\Omega_\mathrm{p}}{0.2} \right)^{-1}\left(\frac{\ln N}{25}\right),
\end{align}
where $T_\mathrm{CR} = 2\pi/\pattern$ is the orbital period at corotation.
Thus, if it has to exponentiate out of Poisson noise, a rather vigorously growing spiral with $\beta/\pattern \gtrsim 0.1$ will saturate after a few Gyr.
More realistic galactic noise tends to be of higher amplitude than is implied by Poisson statistics \citep{sellwood2012spiral,Fouvry2015-nk}, though this will only change the timescale $t_\mathrm{sat}$ by a logarithmic factor.

A caveat to our work is that we have considered a highly idealized system in which we isolated the interaction of a single spiral mode with an ambient stellar population. In reality, there can be multiple modes at play simultaneously.
Strong coupling between different modes can cause spirals to saturate \citep{Sygnet1988-eq,laughlin1997spiral}, although \cite{sellwood2022spiral} found that mode-coupling was not a significant effect in their simulations. Still,
superposing many additional weak modes on top of our primary one will have some diffusive effect on stellar orbits, as will as other noise sources like molecular clouds.  Moreover, spiral waves can be damped by hydrodynamic interactions with the interstellar gas \citep{kalnajs1972damping,roberts1972role,dekker1976spiral}. One can get an idea of the impact these effects will have on our primary spiral mode by characterising them all, very crudely, by a simple `dissipation timescale' $\nu^{-1}$.  Then in the linear regime, the spiral growth rate $\beta$ will be replaced by $\beta - \nu$, meaning that if $\nu \gtrsim \beta$ the mode in question will never grow out of the noise.  In the more realistic case that $\nu \ll \beta$, the linear physics will proceed essentially unchanged from the case $\nu=0$.
but as the mode amplitude approaches the nonlinear regime, one effect of the additional forces will be to de-trap stellar orbits \citep{johnston1971dominant,auerbach1977collisional,Hamilton2023}.  This will counteract the flattening of the DF near corotation and hence allow the spiral amplitude to continue to grow.  Thus, we speculate that more noisy/gas rich galaxies may allow for somewhat higher amplitude spirals.

\section{Summary}
\label{sec:Summary}

A complete theory of spiral structure should explain not only the origin, but also the amplitude, of spiral waves.  It is well known that, at least in isolated, razor-thin, gas-free stellar disks, spiral structure can arise as a linear instability of the underlying DF.  In this paper we have calculated the saturation amplitude of such an instability under the assumption that transferral of angular momentum between stars and spiral modes primarily occurs at the corotation resonance. We find that (i) saturation occurs when the nonlinear resonant libration frequency of the stars is comparable to the initial linear growth rate of the instability $\beta$, and (ii) for galaxies with flat rotation curves,
the resulting relative saturation amplitude in surface density is $\sim (\beta/\pattern)^2\cot \alpha$, where $\pattern$ is the spiral pattern speed and $\alpha$ is its pitch angle.



\section*{Acknowledgements}

I thank the referee Prof J. Binney for a close reading and insightful report, and Prof J. Sellwood for critiquing the initial preprint.
I also thank M. Weinberg, U. Banik, L. Arzamasskiy and S. Tremaine for comments on a putative version of this calculation, M. Zaldarriaga and T. Yavetz for several useful conversations, and
M. Petersen and V. Duarte
for related discussions about the saturation of instabilities in spherical stellar systems and plasmas respectively.
C.H. is supported by the John N. Bahcall Fellowship Fund at the Institute for Advanced Study.

\section*{Data availability}
No new data were generated or analysed in support of this research.




\bibliographystyle{mnras}
\bibliography{Bibliography} 

\appendix 

\section{Angular momentum deficit of resonantly trapped stars}
In this Appendix we provide mathematical details on the calculation of the angular momentum lost by phase-mixed distribution of stars trapped in an arbitrary resonance.
We thereby derive the angular momentum transfer at corotation, which leads us to equation \eqref{eqn:Delta_L_final}.

Start with an $m-$armed spiral perturbation which rotates rigidly in the $\varphi$ direction with pattern speed $\pattern > 0$.
In general, near locations $\bJ_\mathrm{res}$ in action space such that
 \begin{equation}
     \mathbf{N}\cdot \mathbf{\Omega}(\bJ_\mathrm{res}) = N_\varphi \pattern,
     \label{eqn:resonance_condition}
 \end{equation}
for some vector of integers $\boldsymbol{N} = (N_R, N_\varphi)$, we describe the dynamics by first making a canonical transformation to a new set of coordinates
(\citealt{lichtenberg2013regular,Binney2020-mw,tremaine2023dynamics}).
Precisely, we map
$(\btheta, \bJ) \to (\btheta',\bJ')$, where $\btheta' = (\theta_\mathrm{f}, \theta_{\mathrm{s}})$ consists of the `fast' and `slow' angles
\begin{align}
\theta_{\mathrm{f}} \equiv \theta_R, \,\,\,\,\,\,\,\,\,\,\,\, \theta_s \equiv \mathbf{N}\cdot \boldsymbol{\theta} - N_\varphi \pattern t,
\label{eqn:slow_angle}
\end{align}
and $\bJ' = (J_{\mathrm{f}}, J_\mathrm{s})$
consists of the corresponding fast and slow actions
\begin{align}
J_{\mathrm{f}} \equiv  J_R-L N_R/N_\varphi, \,\,\,\,\,\,\,\,\,\,\,\,
J_\mathrm{s} \equiv L/N_\varphi.
\label{eqn:slow_action}
\end{align}
Having made this transformation we may express the Hamiltonian \eqref{eqn:H} in terms of the new coordinates and then average over the fast angle; the result is
\begin{align}
    \mathcal{H}  &\equiv \frac{1}{2\pi}\int \md \theta_\mathrm{f} \, H(\btheta', \bJ') \nn 
    \\
    &=  H_0(\bJ') - N_\varphi \pattern J_\mathrm{s}
    +
    \sum_{k\neq 0} \Psi_{k}(\bJ') \exp(ik {\theta}_\mathrm{s}),
    \label{eqn:averaged_H}
 \end{align}
 where (unlike in equation \eqref{eqn:Energy_Wave}) we expanded in the fast-slow angles, $\delta \Phi = \sum_{\bk} \exp(i\bk\cdot\btheta')\Psi_{\bk}$, and
 employed the shorthand $\Psi_{(0,k)} = \Psi_k$.

Next, we let 
\begin{equation}
    I \equiv J_\mathrm{s} - J_{\mathrm{s, res}}(J_\mathrm{f}) , \,\,\,\,\,\,\,\,\,\,\, 
    \phi_k = \theta_\mathrm{s} + \frac{\arg \Psi_k}{k},
    \label{eqn:I_phi_definitions}
\end{equation}
where $\phi_k \in (-\pi, \pi)$, and $J_{\mathrm{s, res}}$ is the resonant value of the slow action at fixed $J_\mathrm{f}$.
Expanding $\mathcal{H}$ around the resonance for small $I$, discarding constants and unimportant higher order terms, and assuming that the sum over $k$ is dominated by a single $k$ value, $\mathcal{H}$ reduces to the pendulum form \citep{Chirikov1979-fj,lichtenberg2013regular}:
\begin{equation}
    \mathcal{H} = \frac{1}{2}GI^2 - K\cos k\phi,
    \label{eqn:resonant_Hamiltonian}
\end{equation}
where $\phi =\phi_k$,
\begin{equation}
    G(J_\mathrm{f}) \equiv \frac{\partial^2 H_0}{\partial J_\mathrm{s}^2}\Bigg\vert_{J_{\mathrm{s,res}}}, \,\,\,\,\,\,\,\,\,\, 
    K(J_\mathrm{f}) = -2\vert \Psi_k(J_{\mathrm{s,res}})\vert .
    \label{eqn:F_and_G}
\end{equation}
The variables $(\phi, I)$ are canonical variables for the pendulum Hamiltonian \eqref{eqn:resonant_Hamiltonian}.
The pendulum moves at constant `energy' $\mathcal{H}$,  either on an untrapped `circulating' orbit with $\mathcal{H} > K$ or a trapped `librating'  orbit with $\mathcal{H} < K$ (the separatrix between these two families is $\mathcal{H} = K$).
The libration period for oscillations around $(\phi, I) = (0,0)$ is $ t_\mathrm{lib} \equiv 2\pi/\omega_\mathrm{lib}$ where 
\begin{equation}
    \omega_\mathrm{lib}(J_\mathrm{f}) \equiv \sqrt{kKG}.
    \label{eqn:t_libration}
\end{equation}
The maximum width in $I$ of the librating `island' is at $\phi = 0$, where it spans $I\in (-I_\mathrm{h}, I_\mathrm{h})$ and $I_\mathrm{h}$ is the \textit{island half-width}:
\begin{equation}
    I_\mathrm{h}(J_\mathrm{f}) \equiv 2\sqrt{\frac{K}{G}}.
    \label{eqn:half_width}
\end{equation}
For ease of notation we will mostly drop the $J_\mathrm{f}$ dependence of each quantity from now on.

Consider the behavior of the DF $f(\phi, I, t)$ with initial value $f_0$.  The change to the angular momentum carried by the near-resonant stars at time $t$ compared to time $0$ is 
\begin{align}
    \Delta \mathcal{L}_\mathrm{res}(t) = 2\pi \int& \md J_\mathrm{f} \md \phi\, \md I\, [N_\varphi I + J_\mathrm{s,res}][f(\phi, I, t) - f_0(\phi, I)],
    \label{eqn:Angular_Momentum_Loss}
\end{align}
where we used the transformations \eqref{eqn:slow_action} and \eqref{eqn:I_phi_definitions}.
The term $J_\mathrm{s,res}$ can be dropped since the integral involving this term just expresses the change in the total number of stars, which is zero.
We now use the approximation that the majority of changes in $f$ occur in the librating region of phase space, as opposed to the circulating region.
With this simplification, the only 
contributions to the right hand side of \eqref{eqn:Angular_Momentum_Loss} come from integrating over the libration region. 
Moreover, in steady state we know that the DF must be \textit{phase-mixed} within and around the resonant island, i.e. in the vicinity of the resonance the steady-state DF will depend on $(\phi, I)$ only through the `energy' $\mathcal{H}$ (equation \eqref{eqn:resonant_Hamiltonian}).
It follows that for
librating orbits the steady-state DF has the symmetry property
\begin{equation}
    f(\phi, I, J_\mathrm{f}) = f(\phi, -I, J_\mathrm{f}), \,\,\,\,\,\,\,\,\,\, (\mathrm{librating \,\, orbits,} \,\, t\gg t_\mathrm{lib}).
\end{equation} 
These properties allow us 
to reduce the right hand side of \eqref{eqn:Angular_Momentum_Loss} in steady state to the following form:
\begin{align}
    \Delta \mathcal{L}_\mathrm{res}(t\to\infty) = -2\pi N_\varphi \int& \md J_\mathrm{f}  \int_{-\pi}^\pi \md \phi\, \int_0^{I_+(\phi)}  \md I \, I  \nn 
    \\
    &\times [f_0(\phi, I) - f_0(\phi, -I)],
    \label{eqn:time_asy_DeltaLres}
\end{align}
where $I_+(\phi) = I_\mathrm{h} \cos (k\phi/2)$ is the maximum $I$ value of the librating region at a fixed $\phi$.
Crucially, the integral we must now perform only depends on the initial DF $f_0$, and not the final DF.
Note that equation \eqref{eqn:time_asy_DeltaLres} is very general: it only depends on the assumptions that (i) phase mixing is complete and (ii) there is little or no contribution from circulating orbits. In particular, it
does not depend on the rate at which the perturbation grew (see \citealt{Dewar1973-rf}). 

Further assuming $f_0$ is independent of $\phi$, we can reverse the order of integration and find:
\begin{align}
    \Delta \mathcal{L}_\mathrm{res}
    =
    -8\pi N_\varphi \int \md J_\mathrm{f} 
    \int_0^{I_\mathrm{h}} \md I \,
        I  [f_0(I) - f_0(-I)]\cos^{-1}\frac{I}{I_\mathrm{h}}.
        \label{eqn:Angular_Momentum_f_Differences}
\end{align}
We can calculate this integral explicitly if we expand $f_0$ around $I=0$, which is valid if the resonance is narrow compared to the whole angular momentum space:
\begin{equation}
    f_0(I) - f_0(-I) = 2I f_0'(0) + \mathcal{O}(I^3).
\end{equation}
Keeping only the leading term, equation \eqref{eqn:Angular_Momentum_f_Differences} becomes
\begin{align}
    \Delta \mathcal{L}_\mathrm{res} =  -\frac{32\pi N_\varphi}{9} \int \md J_\mathrm{f} \left[ I_\mathrm{h}^3  \, \pder{f_0}{I }\right]_{J_\mathrm{s} = J_\mathrm{res}(J_\mathrm{f})}.
\label{eqn:Delta_L_almost_final}
\end{align}

If we finally specialise to the corotation resonance of an $m$-armed spiral mode, then $N_\varphi=m$, $N_R=0$, $k=1$, $I = (L-L_\mathrm{CR})/m$, $\phi = m(\theta_\varphi - \pattern t)$, $J_\mathrm{f} = J_R$ and $ \Psi_k  =  \Phi_{m} $.  With these substitutions, equation \eqref{eqn:Delta_L_almost_final} is equal to $\Delta \mathcal{L}_\mathrm{res} = -\mathcal{L}_\mathrm{mode}$ where $\mathcal{L}_\mathrm{mode}$ is given in equation \eqref{eqn:Delta_L_final}.

\label{lastpage}

\end{document}